\documentclass[12pt]{iopart}

\usepackage{graphicx}
\usepackage[utf8]{inputenc}
\usepackage{cite}
\usepackage{microtype}

\usepackage{hyperref}
\hypersetup{
	colorlinks=true,
	linkcolor=blue,
	filecolor=magenta,
	urlcolor=blue,
	citecolor=blue,
}
\usepackage{layouts}

\begin{document}

\title[Anomalous Ising freezing times]{Anomalous Ising freezing times}

\author{J. Denholm and B.\ Hourahine}

\address{SUPA and Department of Physics, University of Strathclyde, 107 Rottenrow East, Glasgow, G4 ONG}

\ead{j.denholm@strath.ac.uk}

\vspace{10pt}

\begin{indented}
\item[] \today
\end{indented}

\begin{abstract}
We measure the relaxation time of a square lattice Ising ferromagnet that is quenched to zero-temperature from supercritical initial conditions. We reveal an anomalous and seemingly overlooked timescale associated with the relaxation to ``frozen'' two-stripe states. While close to a power law of the form $\sim L^{\nu}$, we argue this timescale actually grows as $L^{2}\ln{L}$, with $L$ the linear dimension of the system. We uncover the mechanism behind this scaling form by using a synthetic initial condition that replicates the late time ordering of two-stripe states, and subsequently explain it heuristically.
\end{abstract}

%
%
%
%
%

\section{Introduction}
The zero-temperature coarsening of the 2D Ising ferromagnet was thought fully understood through the dynamical scaling hypothesis~\cite{Bray1993}. From a symmetric and unmagnetised initial state, magnetic domains nucleate and grow in length as the square root of time, before ultimately engulfing the system~\cite{Bray1993, Humayun1991}---however, the dynamics is richer still.

Surprisingly, the ground state is not always reached. Instead, 34\% of realisations ``freeze'' into two-stripe states, which span in only a single lattice dimension and are forever trapped at constant energy~\cite{Spirin2001a, Spirin2001b}. Additionally, the ground state is only reached on a timescale of $\mathcal{O}(L^{2})$ around $62\%$ of the time, with $L$ the linear dimension of the system~\cite{Spirin2001a, Spirin2001b}. Perhaps most interestingly, $4\%$ of the realisations reach ``diagonal'' winding configurations, which dramatically forestall the relaxation process by slowly decaying to homogeneity on a timescale of $\mathcal{O}(L^{3.5})$~\cite{Spirin2001a, Spirin2001b}.

The probability of observing each of these topologically distinct behaviours is seemingly identical to the equivalent crossing probability in continuum percolation, which correspondingly explains how the ``fate'' of the system is sealed~\cite{Barros2009, Olejarz2012, Cardy1992, Pinson1994, Pruessner2004, Langlands1992}. The role of critical percolation in 2D curvature driven coarsening has been studied extensively~\cite{Arenzon2007, Blanchard2013, Blanchard2014, Cugliandolo2016, Blanchard2017, Corberi2017, Mullick2017, Godreche2018, Ricateau2018, Tartaglia2018}.

In one spatial dimension, the ground state is always reached~\cite{krapivsky2010}, and in three-dimensions, the ground state is essentially never reached~\cite{Newman1999, Newman2000, Olejarz2011a, Olejarz2011b}. The final states in three dimensions are a host of non-static topologically complex configurations that remain at iso-energy \emph{ad infinitum}~\cite{Olejarz2011a, Olejarz2011b}.

In this work, we uncover an apparently overlooked timescale associated with the relaxation to ``frozen'' two-stripe states. Our main result is that this anomalous timescale appears to grow as $\sim L^{2}\ln L$---i.e.\ the dynamics is slower than the standard coarsening time of $\sim L^{2}$. We detail the model and method in Sec.~\ref{sec:methods}, and introduce the new timescale in Sec.~\ref{sec:relaxation_timescales}. In Sec.~\ref{sec:anomalous_times} we analyse this new scaling form and explain its origin.

\section{Zero-temperature Ising Model}
\label{sec:methods}
We consider nearest-neighbour interactions between magnetic spins on periodically bounded square lattices of length $L$, which is equivalent to wrapping the square lattice around a torus. The spins are denoted by $S_{i} = \pm 1$ and may point up or down respectively. The total energy of the system is given by the Hamiltonian
\begin{equation}
    H = -J\sum\limits_{i, j} S_{i}S_{j},
\end{equation}
where ${J} > 0$ is a ferromagnetic coupling constant controlling the strength of the nearest-neighbour interactions, $i$ indexes each spin in the system and $j$ the nearest-neighbours of each $S_{i}$.

Operationally we use single spin-flip zero-temperature Glauber dynamics: energy lowering and conserving flips are accepted with probabilities of $1$ and $0.5$ respectively, while energy raising flips are forbidden~\cite{Glauber1963, krapivsky2010}. To mimic infinite temperature initial conditions, we initialise each realisation of the dynamics by randomly ordering a microstate of zero-magnetisation.

We implement the dynamics using a standard continuous-time rejection-free Monte Carlo method~\cite{Bortz1975, Sahni1983, Hassold1993, Landau2009}. Each site has the chance to flip once, on average, in a single Monte Carlo time step. Let the probability that a given site $S_{i}$ should flip be $p_{i}$. The total probability in the system is $P=\sum p_{i}$. To flip a spin, one selects a site with probability $p_{i} / P$ and reverses its orientation~\cite{Bortz1975, Sahni1983, Hassold1993, Landau2009}. Time then advances by $\Delta t = -\ln r / P$, where $r \in (0, 1)$ is a uniform random number and $\left< -\ln r\right> = 1$. This process is repeated until no flippable spins remain. The advantage of this method is that one only visits flippable spins, thus avoiding the simulation of null processes. Intuitively, $\Delta t$ is actually the time between accepted spin flip events in the standard rejection-based Monte Carlo method~\cite{Bortz1975, Sahni1983, Hassold1993, Landau2009}.

\begin{figure}[t!]
    \includegraphics[width=\textwidth]{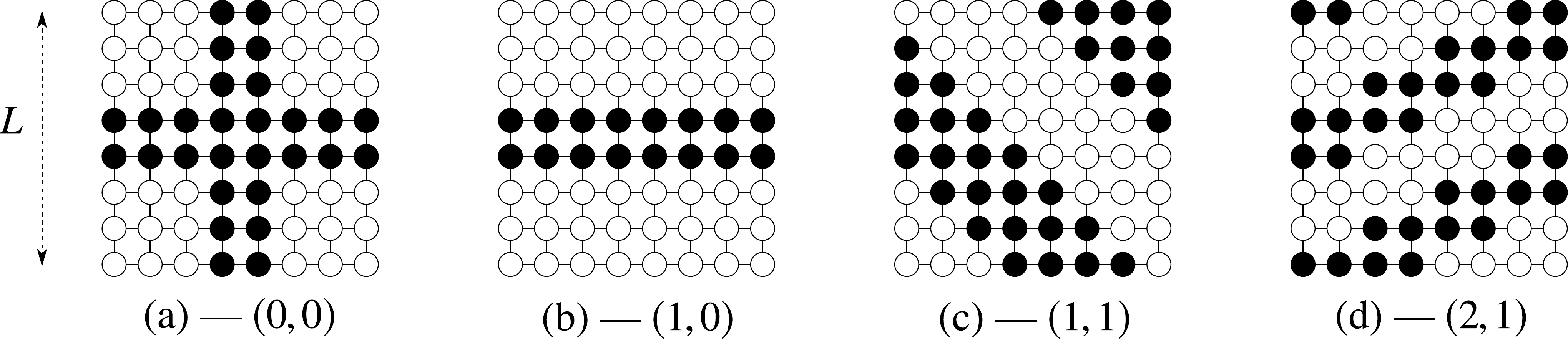}
    \caption{(a) Ground state topology: a \emph{single} domain winds in \emph{both} lattice dimensions. (b) On-axis stripe topology: \emph{both} domains wind in a \emph{single} lattice dimension. (c) Diagonal stripe topology: \emph{both} domains wind in \emph{both} lattice directions. (d) Diagonal stripe topology: \emph{both} domains wind \emph{once} vertically and \emph{twice} horizontally. (c) and (d) are both unstable and ultimately collapse to the ground state.}
    \label{fig:ising_topologies}
\end{figure}

The unique classes of spanning clusters that arise in the zero-temperature Ising model are distinguished by the winding numbers of continuum percolation. A cluster of winding number $(a, b)$ winds $a$ times horizontally (toroidally) and $b$ times vertically (poloidally)---as shown in Fig.~\ref{fig:ising_topologies}. In Fig.~\ref{fig:ising_topologies}~(a), the configuration exhibits the ground state topology, meaning it necessarily collapses to the ground state. In Fig.~\ref{fig:ising_topologies}~(b), the configuration is that of an infinitely long-lived frozen two-stripe state. In Fig~\ref{fig:ising_topologies}~(c)--(d), we see unstable diagonal winding configurations that ultimately reach the ground state.

Clusters of winding numbers $(\pm a, \pm b)$ are equivalent both physically and in probability, so we set $a, b \ge 0$. This is also true of $(1, 0)$ and $(0, 1)$ windings, so we denote all on-axis stripe states as $(1, 0)$.

\section{Overview of timescales}
\label{sec:relaxation_timescales}
The extinction and survival probabilities are useful diagnostics for highlighting the multiple timescales at play, and are defined as follows. Let the extinction time be the time for a given realisation to reach its final state. The distribution of the extinction time is the extinction probability, $E(t)$, which upon integration, gives the probability of reaching a final state within some time interval $\Delta t$.

The survival probability is simply the likelihood that the system is still active at time $t$. Trivially, the survival probability is
\begin{equation}
S(t) = 1 - \int\limits_{0}^{t} E(t)dt.
\label{eqn:survival_prob}
\end{equation}
The cumulative of $E(t)$ gives the fraction of realisations that have reached their final state by time $t$, and the remaining fraction, or the survival probability, is simply one minus the cumulative. Note, Eqn.~\ref{eqn:survival_prob} assumes $E(t)$ is a normalised probability density function.

For each of the distinct timescales, there is an associated relaxation time $\tau(L)$~\cite{Spirin2001a, Olejarz2012}. Realisations reaching their final state throughout this time cause the survival probability to undergo an exponential decay of the form $\sim \exp(-t/\tau(L))$~\cite{Spirin2001a, Olejarz2012}. The two known timescales in the zero-temperature Ising model should cause two exponential decay regimes in $S(t)$, with decay constants $\tau(L) \sim L^{2}$ and $\tau(L) \sim L^{3.5}$~\cite{Spirin2001a, Olejarz2012}. On a semi-logarithmic scale, these appear as linear decays with slopes of $-1 / \tau(L)$.

\begin{figure}[t!]
	\centering
    \includegraphics[width=0.75\columnwidth]{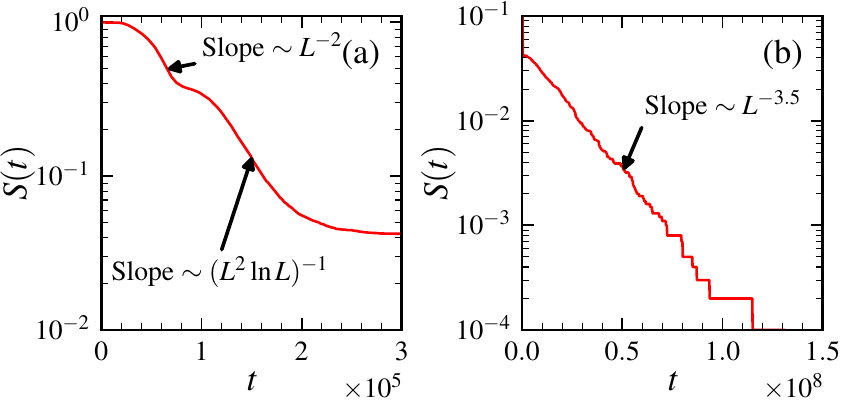}
    \caption{(a) Survival probability versus time for  $0 \le t \le 3 \times 10^{5}$. (b) Slow decaying tail of $S(t)$ corresponding to the relaxation of diagonal stripe states. The data are based on $10^{4}$ realisations with $L=512$.}
    \label{fig:survival}
\end{figure}

Consider now the survival probability in Fig.~\ref{fig:survival}; although we expected two decay regimes, there are actually \emph{three}. Firstly, we see the standard ground-state coarsening timescale of $\mathcal{O}(L^{2})$ (Fig.~\ref{fig:survival}~(a)). Secondly, we see an additional timescale that grows as $\sim L^{2} \ln L$, which is actually associated with realisations reaching frozen two-stripe states (Fig.~\ref{fig:survival}~(a)). Finally, we see the large timescale of $\mathcal{O}(L^{3.5})$ discovered by Spirin \emph{et al.}~\cite{Spirin2001a}, which causes the slow decay in Fig.~\ref{fig:survival}~(b). If one examines the survival probabilities presented in Refs.~\cite{Spirin2001a, Olejarz2012}, there are subtle hints of this intermediate timescale, but its influence on $S(t)$ is almost imperceptible in the small system sizes used.

\subsection{Survival analysis}

\begin{figure}[h!]
    \centering
    \includegraphics[width=0.75\columnwidth]{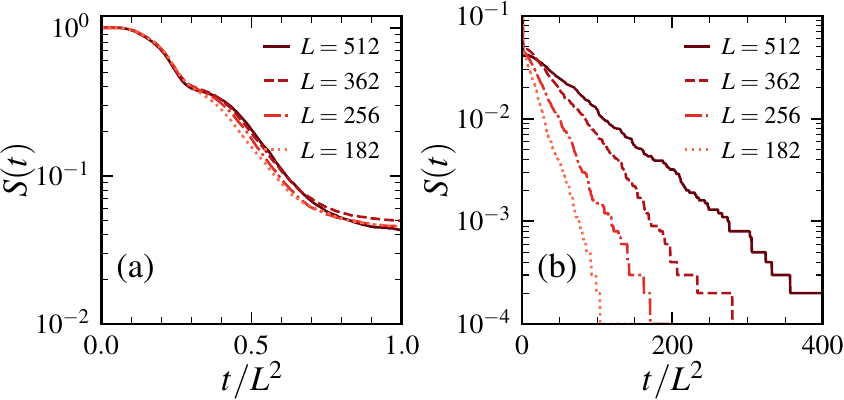}
    \caption{Survival probability versus $t / L^{2}$ for (a) $0 \le t / L^{2} \le 1$ and (b) $0 \le t / L^{2} \le 400$ from systems of length $L$. The data are based on $10^{4}$ realisations.}
    \label{fig:multi-L-survival}
\end{figure}

With these three decay regimes in mind, we now examine the survival probability as a function of $t / L^{2}$ for specified $L$ in Fig.~\ref{fig:multi-L-survival}. In the first decay regime---which is caused by the ground-state timescale of order $L^{2}$---we see data collapse in $S(t)$ (see Fig.~\ref{fig:multi-L-survival}~(a)). However, in the second decay regime---which is caused by the anomalous timescale of order $L^{2}\ln L$---the data collapse fails (see Fig.~\ref{fig:multi-L-survival}~(a)). Finally, we see the long tails caused by the slow relaxation of diagonal-stripe states in Fig.~\ref{fig:multi-L-survival}~(b).

\begin{figure}[h!]
    \centering
    \includegraphics[width=0.75\columnwidth]{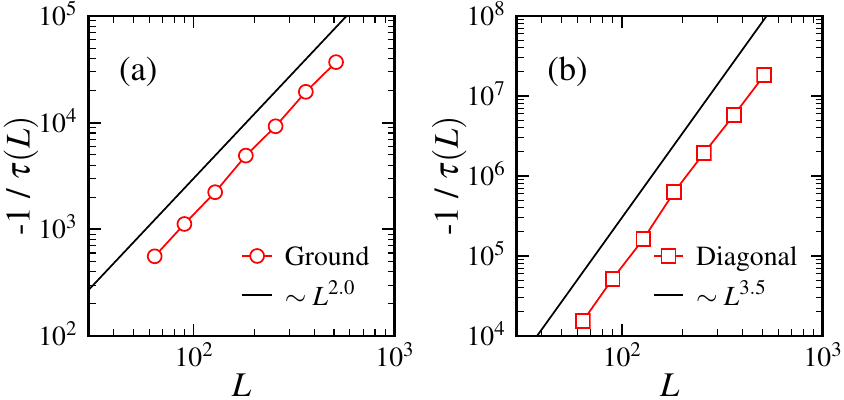}
    \caption{Estimates of the slopes in $\ln S(t)$ versus $t$ over (a) the ground-state decay and (b) the diagonal-stripe state decay versus $L$.}
    \label{fig:survival_slopes}
\end{figure}

Since we know the form of the survival probability, we can actually recover the slopes quoted in Fig.~\ref{fig:survival} by fitting straight lines to $\ln S(t)$ versus $t$ over each region and computing how the slope scales with $L$. Distinguishing between the $L^{2}$ and $L^{2}\ln L$ timescales is difficult due to their overlap---particularly at small $L$---so we apply the analysis to the $L^{2}$ and $L^{3.5}$ timescales only. We plot these estimates in Fig.~\ref{fig:survival_slopes}.
It is important to note that these estimates are necessarily crude because one must first judge over which region to apply the fit.

\subsection{Synthetic initial conditions}
\begin{figure}[b!]
    \centering
    \includegraphics[width=0.5\columnwidth]{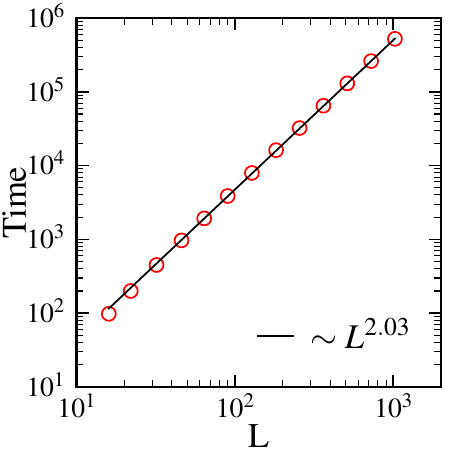}
    \caption{(a) Ground state freezing time from the cross initial condition in Fig.~\ref{fig:ising_topologies}~(a) with a power law fit. The data are based on $10^{4}$ realisations.}
    \label{fig:ground-state-time}
\end{figure}
Before we turn our focus to the anomalous timescale, we recover the known timescales from our simulations using the synthetic initial conditions shown in Fig.~\ref{fig:ising_topologies}. These conveniently allow us to select the topology of the evolution we wish to study. We begin with the ground state freezing time, which grows with the system size as $L^{2}$. We demonstrate this using the cross initial condition in Fig.~\ref{fig:ising_topologies}~(a)---the result of which we plot in Fig.~\ref{fig:ground-state-time}. As $L$ increases, the data exhibit a subtle downward curvature that is indicative of finite size effects.

\begin{figure}[h!]
    \centering
    \includegraphics[width=0.8\columnwidth]{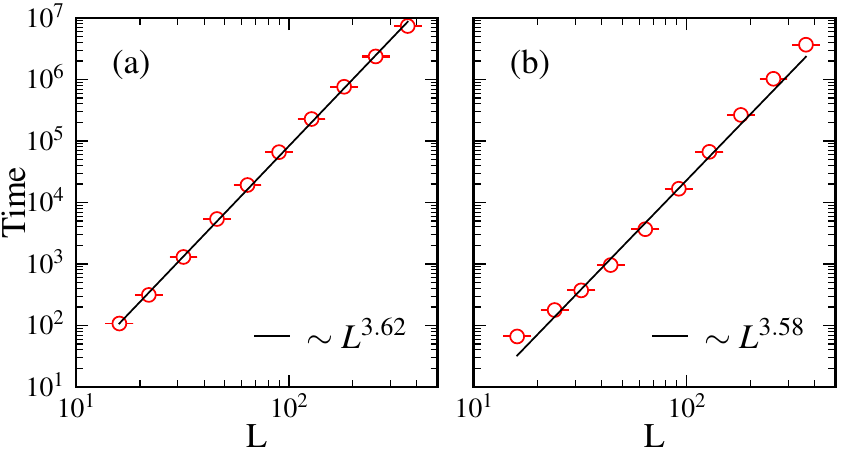}
    \caption{Diagonal stripe state relaxation times versus $L$ from the (a) $(1, 1)$ and (b) $(2, 1)$ initial conditions shown in Fig.~\ref{fig:ising_topologies}~(c)--(d). The data are based on $10^{4}$ realisations.}
    \label{fig:diagonal-freezing-times}
\end{figure}

Spirin \emph{et al.} found the relaxation time of the diagonal stripe states to scale as roughly $\sim L^{3.5}$~\cite{Spirin2001a}. We confirm this by realising the dynamics from the $(1, 1)$ and $(2, 1)$ initial conditions shown in Fig.~\ref{fig:ising_topologies}~(c)--(d). We plot the resulting times in Fig.~\ref{fig:diagonal-freezing-times}, and obtain power law fits of exponents $\nu = 3.62$ and $\nu = 3.58$, as expected. In both cases, there is a non-monotonic curvature in the data that suggests these scaling estimates are pre-asymptotic.

\section{Anomalous timescale}
\label{sec:anomalous_times}
Here, we show that the relaxation time of frozen two-stripe states grows as $\sim L^{2}\ln L$. Our argument is based on three key features: (i) The scaling of the relaxation times using supercritical initial conditions; (ii) the scaling using a synthetic ``wedding cake'' initial condition; (iii) an argument based on annihilating random walkers.

\subsection{Supercritical initial conditions}
\label{sec:supercritical_times}
\begin{figure}[t!]
    \centering
    \includegraphics[width=0.7\columnwidth]{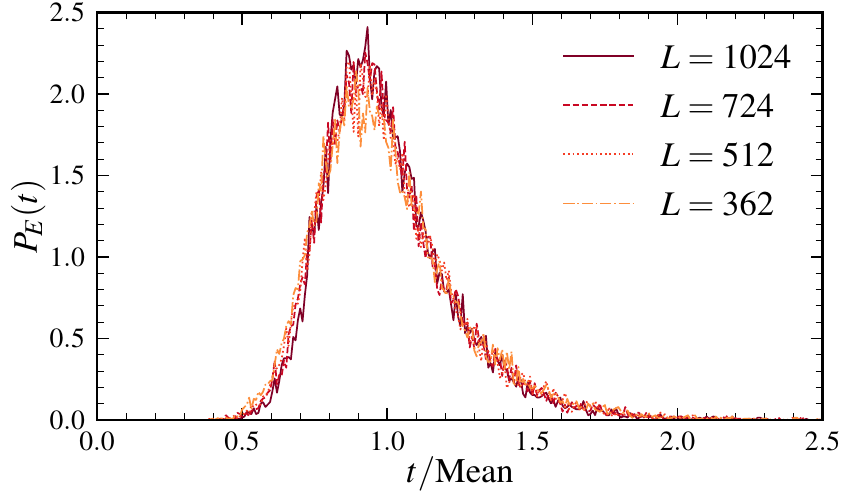}
    \caption{Data collapse in extinction probability for two-stripe states. The distributions are based on two two-stripe state times from ensembles of $5\times 10^{4}$ realisations.}
    \label{fig:stripe_dists}
\end{figure}

We investigate the time scaling of the two-stripe states by realising the dynamics in systems of $16 \le L \le 1024$ from random initial conditions, storing only the subset of times from instances that reach frozen two-stripe states. To test if this relaxation is governed by a single timescale, we consider the distribution of the two-stripe freezing times (see Fig.~\ref{fig:stripe_dists}). In each case, time is rescaled by the mean of the distribution. The data collapse in Fig.~\ref{fig:stripe_dists} indicates the relaxation is governed by a \emph{single} timescale.

\begin{figure}[b!]
    \includegraphics[width=\columnwidth]{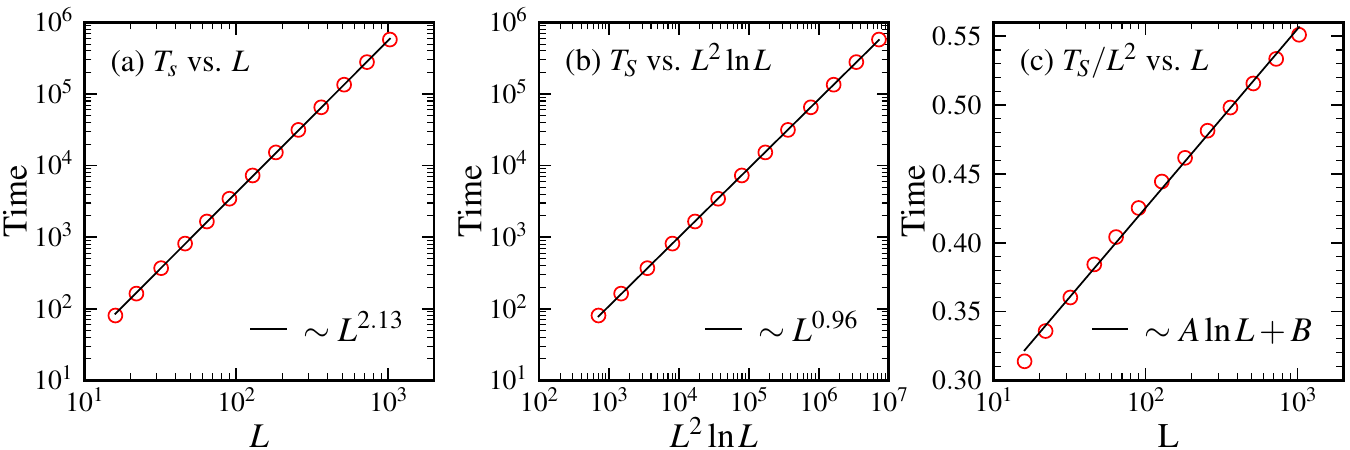}
    \caption{(a) Two-stripe state freezing time $T_{S}$ versus system size $L$. (b) $T_{s}$ versus $L^{2}\ln L$. (c) $T_{S} / L^{2}$ versus $L$ (semi-logarithmic scale). The data are based on the two-stripe states times from ensembles of $5\times10^{4}$ realisations.}
    \label{fig:stripe_times_fits}
\end{figure}

We plot the mean two-stripe state freezing time $T_{S}$ in Fig.~\ref{fig:stripe_times_fits}~(a) as a function of $L$ and obtain a power law fit (solid line) of exponent $\nu=2.13$. Clearly, $T_{S}$ grows faster than $\sim L^{2}$, but the form is not necessarily a simple power law; the near, but greater than integer exponent, along with subtle downward curvature in the data, is evidence of a logarithmic factor.

There are two strong pieces of evidence for $L^{2}\ln L$ scaling. The first is shown in Fig.~\ref{fig:stripe_times_fits}~(b), where we plot $T_{S}$ versus $L^{2}\ln L$---the result of which is linear. A power law fit to these data gives an exponent of $\nu = 0.96$, which is strikingly close to linear considering the presence of finite size effects. The second piece of evidence is the recovery of the logarithm in $T_{S}$: in Fig.~\ref{fig:stripe_times_fits}~(c) we plot $T_{S} / L^{2}$ versus $\ln L$ on a semi-logarithmic scale, and find that even though there is curvature due to finite size effects, there is still discernible evidence of a logarithm. The recovery of $\ln L$ from $T_{S}$ is an important piece of evidence because it is the weakest part of the scaling form, and is discernible even in the presence of finite size effects.

Frozen two-stripe states only occur with probability $\approx 0.34$, so to study them one essentially discards $66\%$ of their simulations.  To reduce the role of finite size effects in our analysis and increase the number of realisations used in our scaling estimates, we seek to study this process more efficiently.

\subsection{Synthetic initial condition}
\label{sec:wedding_cake_times}
\begin{figure}[h!]
    \centering
    \includegraphics[width=0.3\columnwidth]{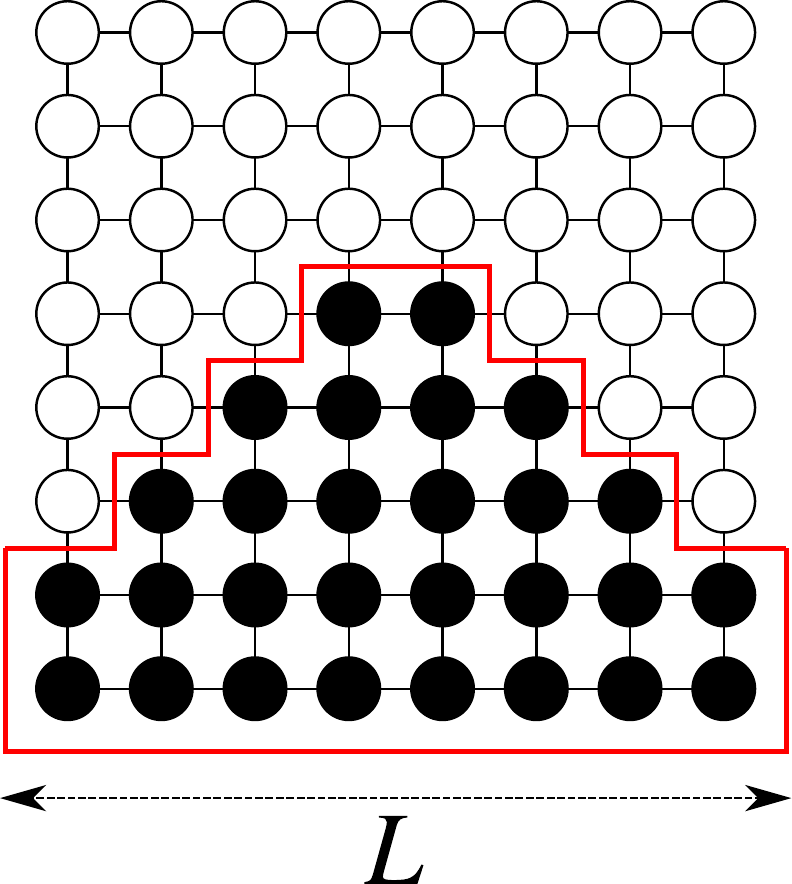}
    \caption{Maximally tiered ``wedding cake'' initial condition.}
    \label{fig:wedding_cake}
\end{figure}

To further explore the two-stripe state relaxation, we use the synthetic ``wedding cake'' initial condition (Fig.~\ref{fig:wedding_cake}).
Realising the dynamics from this configuration is more practical as the required CPU time is lesser, and 100\% of the realisations are used. We can therefore obtain $10^{5}$ realisations with $16 \le L \le 2048$ using the wedding cake initial condition.
\begin{figure}[b!]
    \includegraphics[width=\columnwidth]{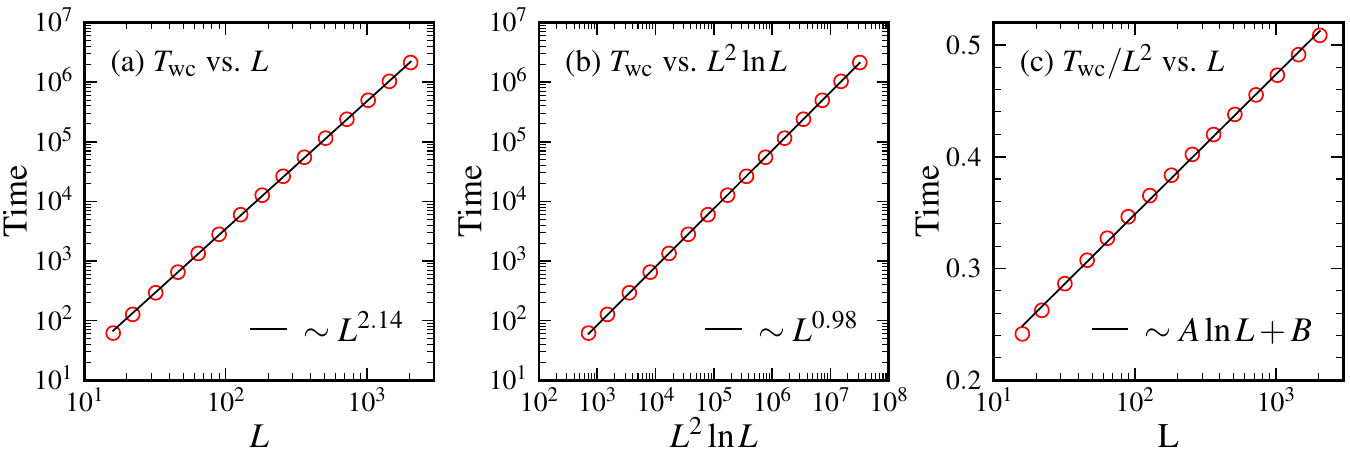}
    \caption{(a) Wedding cake freezing time $T_{\mathrm{wc}}$ versus system size $L$. (b) $T_{\mathrm{wc}}$ versus $L^{2}\ln L$. (c) $T_{\mathrm{wc}} / L^{2}$ versus $L$ (semi-logarithm scale). The data are based on $10^{5}$ realisations from the wedding cake initial condition in Fig.~\ref{fig:wedding_cake}.}
    \label{fig:wedding_cake_fits}
\end{figure}

We perform the same analysis as in Sec.~\ref{sec:supercritical_times} to show the wedding cake relaxation scales as $L^{2}\ln L$. Firstly, we plot the wedding cake freezing time $T_{\mathrm{wc}}$ versus $L$ in Fig.~\ref{fig:wedding_cake_fits}~(a) and obtain a power law fit of exponent $\nu = 2.14$. We again see subtle downward curvature in the data, and the exponent is congruent with that of Fig.~\ref{fig:stripe_times_fits}~(a). Secondly, we plot $T_{\mathrm{wc}}$ versus $L^{2}\ln L$ in Fig.~\ref{fig:wedding_cake_fits}~(b) and obtain a power law fit of exponent $0.98$, thus showing $T_{\mathrm{wc}}$ is linear with $L^{2}\ln L$. Thirdly, we recover the logarithm in $T_{\mathrm{wc}}$ by dividing out $L^{2}$ in Fig.~\ref{fig:wedding_cake_fits}~(c). On the semi-logarithmic scale, $T_{\mathrm{wc}} / L^{2}$ is strikingly linear---especially at large $L$---given that the form is still influenced by finite size effects.

Our analysis using the wedding cake configuration is cleaner compared to the case of supercritical initial conditions as we obtain a greater number of realisations per system size and are able to simulate larger systems.

\subsection{Heuristic argument}
\label{sec:argument}
Our simulations show $L^{2}\ln L$ scaling, so we now seek a theoretical argument to solidify this claim. Here, we use the wedding cake as a simplistic model to probe the origin of the anomalous timescale.

\begin{figure}[b!]
    \centering
	\includegraphics[width=0.75\columnwidth]{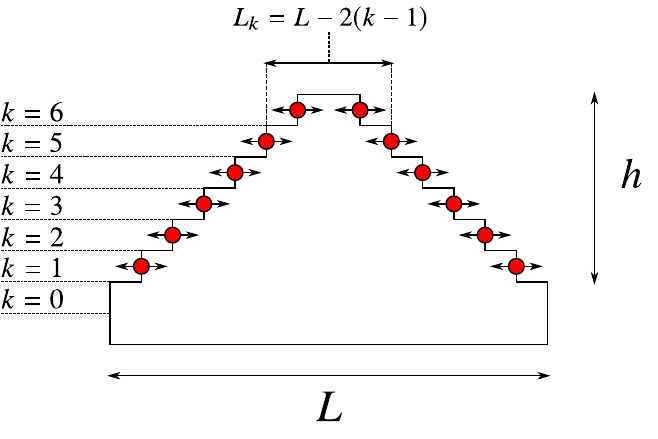}
	\caption{Random walker picture of the ``wedding cake'' initial condition. The red circles cartoon misaligned spin pairs (domain-wall particles) as random walkers that may hop left or right with equal probability, and are reflected by the edges of their tiers.}
	\label{fig:random_walkers}
\end{figure}

First, we cartoon the relaxation of the wedding cake. In the initial condition, each tier has two misaligned spin pairs forming domain-wall particles that behave as random walkers (see Fig.~\ref{fig:random_walkers}).
The walkers may hop to the left or right with equal probability, and and reflected by the boundaries of their tiers. When two walkers on the same tier meet: they annihilate---at which point the tier has relaxed and the number of active sites has reduced by four. The width of tier $k$ is $L_{k} = L - 2(k - 1)$, and the height of the wedding cake is $h = L/2 - 1$ (see Fig.~\ref{fig:random_walkers}).

The walkers on tier $k=5$ are pinned between the positions of the walkers on tiers $k = 4$ and $k = 6$, so they cannot meet. However, the walkers on the topmost tier are unconstrained and therefore free to meet, so the entire structure relaxes ``top-down''.

Each of the tiers in the wedding cake can be modelled as annihilating random walkers on intervals of length $L_{k}$. As the system relaxes from the top down, we assume the relaxation of the topmost tier occurs independently of the other tiers in the system. This is however a quasi-static approximation, because the location of the left- and right-hand boundaries of the topmost tier depends upon the positions of the walkers on the second top-most tier, which do fluctuate.

\begin{figure}[b!]
    \centering
    \includegraphics[width=0.625\textwidth]{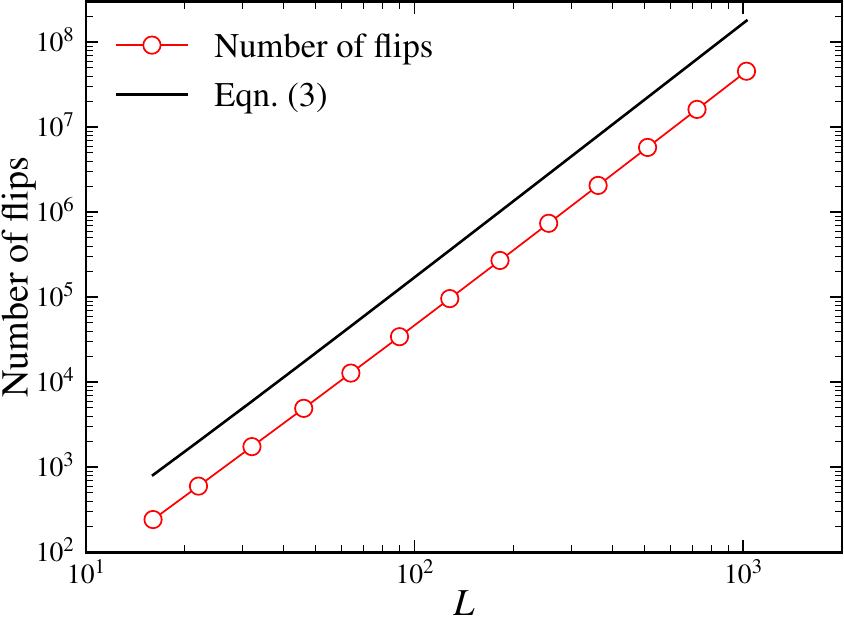}
    \caption{Number of flips in the wedding cake relaxation versus $L$ (circles) and plot of Eqn.~\ref{eqn:number-hops} (line). The data are based on $10^{4}$ realisations.}
    \label{fig:flip-count}
\end{figure}

When a spin is flipped, the corresponding walker hops to the left or right. The \emph{number of hops} required for two random walkers on an interval of length $L_{k}$ to collide is of order $L_{k}^{2}$, ergo the total number of hops required for the entire structure to relax is
\begin{equation}
    N_{\mathrm{H}} \sim \sum_{k=1}^{k=h} L_{k}^{2}
    \label{eqn:number-hops}
\end{equation}
If our argument is to accurately describe the wedding cake relaxation, we should be able to count the number of spin flips in our simulations and find the same scaling as Eqn.~\ref{eqn:number-hops}. We show these quantities in Fig.~\ref{fig:flip-count}, where one sees the forms exhibit equivalent scaling. Eqn.~\ref{eqn:number-hops} only gives the order of the number hops, which explains why the forms in Figure~\ref{fig:flip-count} have the same slopes but different magnitudes.
The congruence between the scaling of the number of hops required for the system to relax and Eqn.~\ref{eqn:number-hops} in Fig.~\ref{fig:flip-count} validates our treatment of each tier as independent random walkers.

Now that we have established the number of flips required for the structure to relax, we seek to relate this quantity to simulation time, which has units of Monte Carlo time steps. Each spin in the system undergoes one microtrial per time step; that is to say, each site has the chance to flip once, on average, in a single time step. A successful microtrial is one in which the spin flip is accepted. As time progresses during a zero-temperature quench, the number of flippable spins decreases, so fewer and fewer spins flip in a given time step. Consequently, the time between successful microtrials---that is, the time between successful spin flips---increases. The mean \emph{time between} successful microtrials is $\Delta t = 1 / P$~\cite{Bortz1975, Sahni1983, Hassold1993, Landau2009}. $\Delta t$ is inversely proportional to the number of flippable spins in the system, which in the wedding cake is $4k$, so we denote the time increment as $\Delta t_{k}$~\cite{Bortz1975, Sahni1983, Hassold1993, Landau2009}.

We know the number of flips required for the tiers to relax, and we have an expression for the time between individual flips, so we can write the total relaxation time as
\begin{equation}
    t = \sum_{k = 1}^{k=h}L_{k}^{2}\Delta t_{k},
    \label{eqn:time1}
\end{equation}
i.e.\ the relaxation time is the sum of the total number of hops (successful flips) multiplied by the time step between each hop. As we use zero-temperature Glauber dynamics, the probability that a given \emph{active site} in the wedding cake configuration should flip---which is an energy conserving move---is $0.5$, except in the very rare exception where two walkers collide, which is energy lowering. Thus, $\Delta t_{k} = 1/(0.5 \times 4k)$, and Eqn.~\ref{eqn:time1} becomes
\begin{eqnarray}
t &\sim& \frac{1}{2}\sum_{k = 1}^{k=h}\frac{1}{k}L_{k}^{2},\label{eqn:sum}\\
  &\sim& \frac{1}{2} \sum_{k = 1}^{k=h} \frac{L^{2}}{k} + \frac{4L + 4}{k} - 4L + 4k - 8. \nonumber
\end{eqnarray}
This form is dominated by the first term in the sum, which asymptotically is
\begin{equation}
t \simeq L^{2}\ln L.
\end{equation}
Although this argument \emph{is} crude, in that it does not completely encapsulate the relaxation process, it is a compelling justification for $ L^{2} \ln L$ scaling.

\section{Discussion \& Conclusion}
\label{sec:discussion}
We identified a new timescale in the zero-temperature coarsening of the square lattice Ising ferromagnet that grew as $L^{2}\ln L$ and arose from the relaxation to ``frozen'' two-stripe states. Our argument for $L^{2}\ln L$ scaling was based on three key features: (i) the time scaling using supercritical initial conditions; (ii) the scaling using a synthetic ``wedding cake'' initial condition; (iii) an argument based on annihilating random walkers.

Using both natural and synthetic initial conditions, we showed that the relaxation times were linear with $L^{2}\ln L$, and by dividing out $L^{2}$ we were able to recover the logarithm. The latter is an important piece of evidence as the logarithm is the most subtle part of the form, and after dividing out $L^{2}$ it was clearly discernible---even in the presence of finite size effects.

We also recovered the known timescales of $\sim L^{2}$ and $\sim L^{3.5}$ as a test of our simulations. Finite size effects hindered our estimates of the exponents in each case; even the simple ground state relaxation, finite size influences caused the \emph{correct} scaling form to yield a \emph{poor fit} at small $L$.

This overlooked timescale is seemingly the same as the relaxation-time of so-called three-hexagon states in the triangular lattice Potts model, and is likely a general feature of edge ordering in kinetic spin systems~\cite{Denholm2019}. One question raised by this work is that of the relaxation timescales in the zero-temperature Potts model, which are markedly different on the square and triangular lattices~\cite{Olejarz2013, Denholm2019}. The relaxation timescales in the square-lattice Potts model are non-trivial and not yet well understood~\cite{Olejarz2013}.

\section*{Acknowledgements}
J.D.\ thanks Sid Redner for invaluable advice and direction throughout this work. The authors thank Leticia Cugliandolo for providing insightful criticisms and suggestions regarding this manuscript, and Gian-Luca Oppo for helpful comments. J.D.\ acknowledges EPSRC DTA5 grant EP/N509760/1 for financial support. The authors acknowledge ARCHIE-WeSt High Performance Computer based at the University of Strathclyde as well as grant EP/P015719/1 for computer resources.

\bigskip
\bibliographystyle{ieeetr}
\bibliography{references}

\end{document}